\documentclass[a4paper,fleqn]{cas-dc}
\usepackage{soul}
\usepackage[sort&compress,numbers]{natbib}
\usepackage{lineno}
\usepackage{setspace}
\usepackage{verbatim}
\usepackage{graphics}
\usepackage{graphicx}
\usepackage{cuted}
\usepackage{lipsum}
\usepackage{multirow}
\usepackage[version=4]{mhchem}
\usepackage{chemformula}

\def\tsc#1{\csdef{#1}{\textsc{\lowercase{#1}}\xspace}}
\tsc{WGM}
\tsc{QE}
\tsc{EP}
\tsc{PMS}
\tsc{BEC}
\tsc{DE}

\begin{document}

\let\WriteBookmarks\relax
\def\floatpagepagefraction{1}
\def\textpagefraction{.001}
\shorttitle{Stacking-Tunable Electronic Properties in HsGDY}
\shortauthors{Fabris \textit{et~al}.}

\title [mode = title]{Stacking-Tunable Electronic Properties in Recently Synthesized Hydrogen-Substituted Graphdiyne}

\author[1]{ Guilherme S. L. Fabris}
\author[2]{ Raphael B. de Oliveira}
\author[1]{ Bruno Ipaves}
\author[3]{ Marcelo L. Pereira Junior}
\author[1]{ Douglas S. Galvão}

\address[1]{Applied Physics Department and Center for Computational Engineering \& Sciences, State University of Campinas, Campinas, São Paulo 13083970, Brazil.}
\address[2]{Department of Materials Science and NanoEngineering, Rice University, Houston, TX 77005, USA.}
\address[3]{University of Brasília, College of Technology, Department of Electrical Engineering, 70910-900, Brasília, Federal District, Brazil.}


\begin{abstract}
Recent progress in porous carbon materials has highlighted the importance of structural design in controlling emergent physicochemical properties. In this context, hydrogen-substituted graphdiyne (HsGDY), a three-dimensional framework derived from graphdiyne (GDY), has recently emerged as a promising architecture whose stacking-dependent behavior remains largely unexplored. Here, we present a comprehensive first-principles investigation of the structural, electronic, and optical properties of HsGDY across distinct stacking sequences. Our results identify the AA and ABC configurations as the most energetically favorable, with AA corresponding to the global minimum, consistent with recent experimental observations. Electronic-structure analysis reveals that HsGDY is an indirect semiconductor with an electronic band gap of 0.89 eV (optB88-vdW), primarily governed by interlayer coupling and van der Waals interactions. The optical response exhibits pronounced absorption features spanning the visible to ultraviolet regions, highlighting strong potential for optoelectronic applications. \textit{Ab initio} molecular dynamics (AIMD) simulations at 700 K confirm the thermal robustness of the framework, with negligible structural distortions. Collectively, these findings elucidate the stacking-dependent stability and semiconducting character of HsGDY, providing a solid theoretical foundation for its integration into next-generation nanoelectronic and energy-harvesting technologies.
\end{abstract}

%\includegraphics[width = \linewidth]{abs-graph}

%\item H1; 
%\item H2;  
%\item H3.


%\linenumbers
\maketitle
\doublespacing

\section*{Introduction}

Recent advances in carbon-based nanomaterials have underscored the central role of hybridization-driven structural design in tailoring physicochemical properties for next-generation nanoelectronics, energy-storage, and optoelectronics technologies \cite{geim2007rise, Peng2014, paupitz2026concise}. The coexistence of sp, sp\textsuperscript{2}, and sp\textsuperscript{3} hybridization states enables the formation of allotropes with markedly distinct structural and functional properties, positioning carbon at the forefront of modern materials science. Following the discovery of graphene in 2004 \cite{novoselov2004electric}, extensive efforts have been devoted to identifying other carbon allotropes capable of overcoming graphene's intrinsic zero band gap, which limits its applicability in semiconductor devices requiring high on-off current ratios \cite{Peng2014, paupitz2026concise}, but retaining some of graphene's properties.

Graphynes (GYs) and graphdiynes (GDYs) constitute an important class of two-dimensional (2D) carbon allotropes formed by replacing C-C bonds in graphene with acetylenic (--C$\equiv$C--) or diacetylenic (--C$\equiv$C--C$\equiv$C--) linkages, yielding planar networks composed of mixed sp and sp\textsuperscript{2} hybridized carbon atoms \cite{paupitz2026concise}. Graphyne was theoretically proposed by Baughman and co-workers in 1987 as a family of crystalline carbon allotropes with high sp content \cite{Peng2014, baughman1987structure}. Although extended graphyne sheets remain experimentally challenging to synthesize \cite{li2018synthesis,RodionovJACS2022,RayPNAS2025}, $\alpha$-graphdiyne ($\alpha$-GDY) was successfully fabricated in 2010 through \textit{in situ} cross-coupling of hexaethynylbenzene monomers on copper foil \cite{paupitz2026concise, li2010architecture, Huang2018}, marking the transition from theoretical predictions to experimentally accessible porous carbon frameworks. Other graphyne-based structures, such as nanotubes \cite{coluci2003families,coluci2004new} and scrolls \cite{Graphyne_NanoScrolls} are also possible.

$\alpha$-GDY exhibits remarkable properties arising from its sp-sp\textsuperscript{2} conjugated structure with uniformly distributed intrinsic nanopores with diameters of approximately 5-6 \AA. First-principles calculations predict a direct electronic band gap of approximately 0.44-1.47 eV for a monolayer, offering a significant advantage over graphene \cite{Huang2018, long2011electronic, zheng2023two}. The electronic band gap can be further tuned through strain, chemical doping, nanoribbon engineering, and functionalization \cite{Huang2018, zheng2023two, Zhao2022}. Moreover, charge carriers may reach mobilities on the order of $10^5$ cm\textsuperscript{2}V\textsuperscript{-1}\textsuperscript{-1}, and the material has been synthesized in multiple morphologies, including nanosheets, nanowires, nanotubes, and three-dimensional (3D) porous structures \cite{Huang2018, Dang2021}. These characteristics have enabled the exploration of graphdiyne-based systems in catalysis, sensing, energy storage, photocatalysis, and biomedical applications \cite{Huang2018, zheng2023two}.

A recent experimental breakthrough is the synthesis of hydrogen-substituted graphdiyne (HsGDY) with a highly ordered AA-stacked structure, achieved via scalable thermal and chemical methods \cite{liu2025aa}. This architecture demonstrates that hydrogen substitution combined with controlled interlayer stacking can significantly enhance lithium adsorption and diffusion through well-defined pore channels, leading to reversible capacities exceeding 1000 mAh g\textsuperscript{-1}. Despite this progress, the stacking-dependent physicochemical properties of HsGDY remain insufficiently understood at the atomistic level, particularly with respect to the interplay between hydrogen functionalization and interlayer registry.

In this work, we have carried out a systematic first-principles investigation of HsGDY. While the experimentally synthesized AA stacking is used as the primary reference, alternative AB and ABC stacking configurations are also examined to evaluate their relative energetic stability and to clarify the influence of interlayer registry on structural, electronic, optical, and mechanical properties. This approach establishes a coherent framework for connecting experimentally observed structures with stacking-dependent trends, providing nanoscale insights into the roles of hydrogen functionalization and interlayer arrangement in this emerging porous carbon material.

\section{Methodology}

The structural stability and electronic and optical properties of HsGDY were investigated using \textit{ab initio} simulations within the framework of Density Functional Theory (DFT) \cite{kohn1965self} as implemented in the \textsc{SIESTA} code \cite{soler2002siesta,garcia2020siesta}. Exchange-correlation effects were described using the nonlocal van der Waals density functional of the optB88-vdW type, also referred to as the vdW/KBM functional \cite{Klime2009}, selected for its demonstrated accuracy in describing dispersion-driven interactions in layered materials. A double-$\zeta$ polarized (DZP) numerical atomic orbital basis set was employed due to its proven reliability in accurately representing carbon-based hybrid frameworks \cite{Bennett2025}. A real-space mesh cutoff of $450$ Ry was adopted together with a $\Gamma$-centered Monkhorst-Pack grid \cite{monkhorst1976special} of $3\times3\times6$ $k$-points for the primitive cell, whereas a $2\times2\times4$ grid was used for stacking calculations. All computational parameters were carefully tested to ensure the convergence of total energies and derived properties.

All structures were fully relaxed until the maximum residual force on each atom was lower than $0.01$ eV/$\r{A}$, and the total energy convergence threshold for the self-consistent field cycle was set to $10^{-6}$ eV. Spin-unpolarized calculations were performed in accordance with the nonmagnetic character of the systems under investigation.

Thermal stability was assessed through \textit{ab initio} molecular dynamics (AIMD) simulations performed in the canonical ensemble using the Velocity-Verlet integration scheme combined with a Nosé-Hoover thermostat \cite{nose-hoover}. Each system was equilibrated for $10$ ps with a time step of $1$ fs at $700$ K, while total energy fluctuations and atomic displacements were continuously monitored throughout the simulations.

Dynamical stability was further evaluated through phonon dispersion calculations based on the density-functional tight-binding (DFTB) approach as implemented in the \textsc{DFTB+} package \cite{Elstner1998, dftb2020, dftb_performance, Hourahine2025}, using an established parametrization together with the \textsc{Phonopy} code \cite{togo2023implementation}. The calculations were performed on a $2\times2\times3$ supercell to ensure an adequate description of the lattice dynamics of bulk HsGDY. This combined computational framework provides a consistent and reliable basis for assessing the structural stability and interpreting the physical properties discussed in the following sections.

\section{Results and Discussion}

This section presents a systematic analysis of the structural and physicochemical properties of hydrogen-substituted graphdiyne, with particular emphasis on the role of stacking configurations in determining its stability and electronic behavior. The discussion is organized to establish a clear structure-property relationship, beginning with the identification of the most favorable stacking arrangements, followed by an assessment of their energetic and dynamical stability. Subsequently, the electronic and optical responses are examined to elucidate how interlayer registry influences the functional characteristics of this layered porous carbon framework.

\subsection{Crystal Structure and Stacking Configurations}

\begin{figure}[pos = b!]
    \centering
    \includegraphics[width=\linewidth]{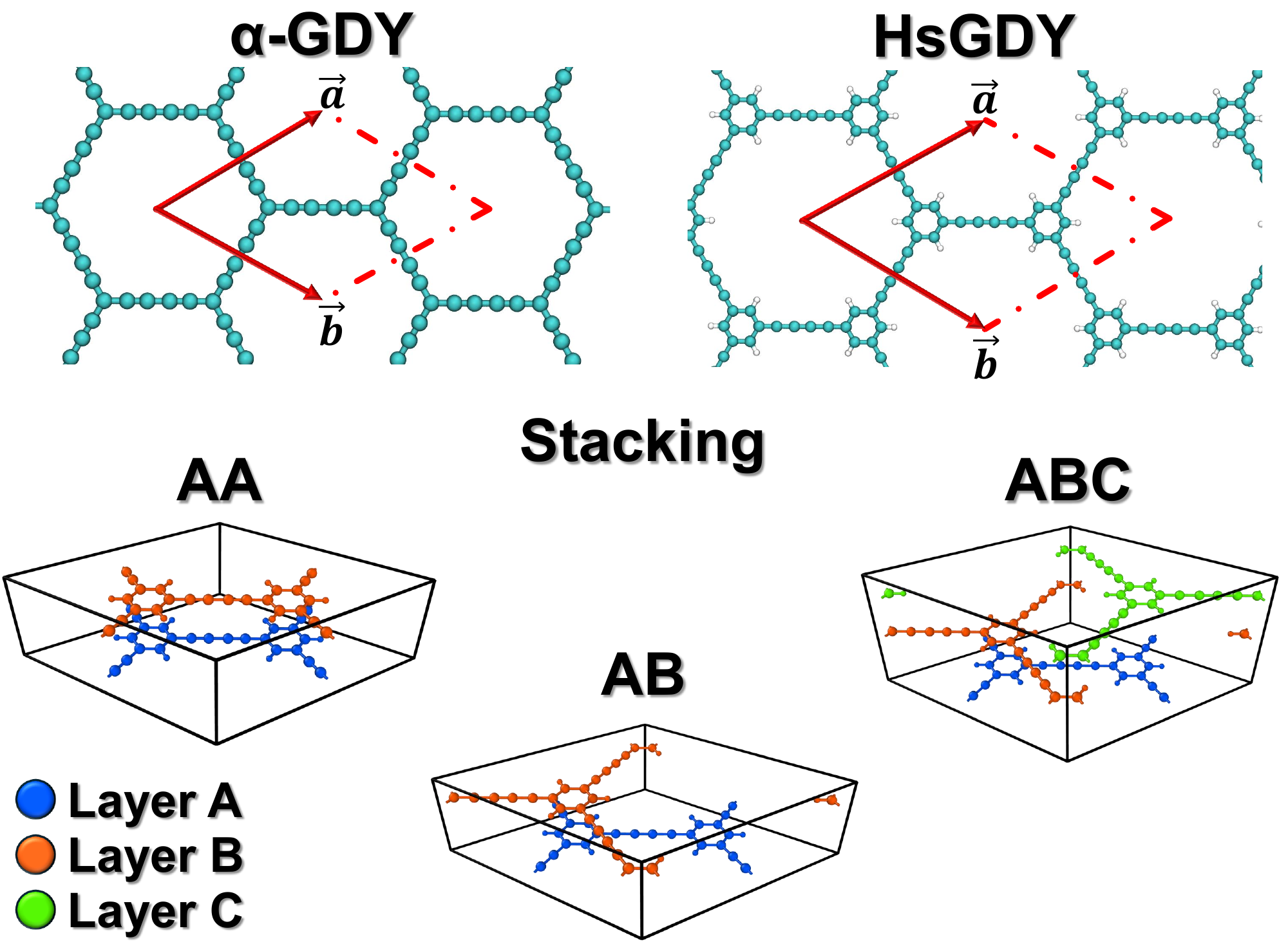}
    \caption{Optimized structures of $\alpha$-GDY and HsGDY, highlighting the hexagonal in-plane lattice defined by the $\vec{a}$ and $\vec{b}$ vectors, together with the AA, AB, and ABC stacking registries. Distinct colors label layers A, B, and C to emphasize the relative lateral offsets that differentiate the stacking sequences.}
    \label{fig:structure-stacking}
\end{figure}

We first establish the structural models of $\alpha$-GDY and HsGDY and define the stacking registries used throughout this work. The layered crystals are built by stacking individual sheets along the out-of-plane direction, in close analogy with graphite, where interlayer cohesion is primarily governed by van der Waals interactions. In addition to the experimentally reported AA arrangement, AB and ABC registries are considered to facilitate a consistent discussion of stacking-dependent trends in the subsequent sections. Figure~\ref{fig:structure-stacking} summarizes the in-plane hexagonal lattices of both materials and illustrates the relative layer registries associated with the AA, AB, and ABC sequences.

For $\alpha$-GDY, structural optimization yields an in-plane hexagonal lattice with $a=b=11.618$ \r{A}, in very good agreement with previous theoretical works on this allotrope \cite{Huang2018, zheng2023two, Niu2013, Li2020}. Within a single layer, the carbon network consists of benzene rings connected through diacetylenic linkages. Along the diacetylene chain, we obtain C-C distances of $1.261$, $1.356$, and $1.261$ \r{A}, while the C-C bond connecting adjacent chains is $1.415$ \r{A}. These values are consistent with the characteristic alternation of $sp$ and $sp^{2}$ bond lengths reported for GDY frameworks \cite{Huang2018, zheng2023two}. Earlier studies of porous GDY sheets similarly report aromatic C-C distances near $1.42$ \r{A}, C=C bonds around $1.34$ \r{A}, and C$\equiv$C bonds close to $1.21$-$1.23$ \r{A} \cite{Huang2018, zheng2023two}.

HsGDY preserves the underlying hexagonal connectivity while incorporating out-of-plane C-H groups at selected aromatic sites, thereby enlarging the in-plane periodicity and modifying the local bonding environment. The primitive cell (one layer) contains six inequivalent atoms, four carbon and two hydrogen atoms, which generate a 30-atom hexagonal unit cell under the symmetry operations of the \textit{P6/mmm} space group. The optimized lattice parameters for the AA-stacked crystal are $a=b=16.63$ \r{A} and $c=6.82$ \r{A}, in close agreement with values inferred from the experimental synthesis \cite{liu2025aa}. The larger in-plane periodicity is reflected in an increased pore diameter of $14.13$ \r{A}. At the hydrogenated aromatic sites, the optimized C-H bond length is $1.106$ \r{A}, consistent with values reported for hydrogenated graphdiyne films and porous graphene derivatives \cite{qiu2018first, lee2014dft}. Aromatic C-C bonds within the rings are slightly elongated to $1.428$ \r{A}, whereas the diacetylene chain exhibits distances of $1.248$, $1.373$, and $1.248$ \r{A}, maintaining the expected alternation of triple and single or double bonds in mixed sp-sp$^{2}$ networks \cite{qiu2018first, lee2014dft}.

The comparison between $\alpha$-GDY and HsGDY indicates that hydrogen substitution perturbs the local geometry without disrupting the global framework. While $\alpha$-GDY remains essentially planar with nearly linear C-C-C angles along the diacetylene segments and an approximately trigonal sp$^{2}$ environment at the benzene rings, HsGDY exhibits small out-of-plane distortions at hydrogenated sites and a broader distribution of C-C-C angles in their vicinity. This behavior is consistent with a localized mixing of sp$^{2}$ and sp$^{3}$ character induced by C-H bonding.

The equilibrium interlayer spacing depends on the stacking registry. After full relaxation, AA, AB, and ABC configurations yield layer separations of $3.411$, $3.273$, and $3.252$ \r{A}, respectively, indicating that the laterally shifted AB and ABC arrangements allow a closer approach between neighboring HsGDY sheets. This trend reflects the balance between interlayer $\pi$-$\pi$ interactions and steric repulsion associated with out-of-plane C-H groups. The lateral offset reduces direct H-H and ring-ring overlap along the vertical direction, enabling a small contraction of the interlayer spacing while preserving the structural integrity of the porous framework.

A direct comparison with the experimental study of Liu \textit{et al.} \cite{liu2025aa} shows that the relaxed AA structure reproduces the measured interlayer spacing of approximately $0.41$ nm obtained from TEM and XRD analyses, and remains consistent with the pore dimensions inferred from BET and diffraction data. Small quantitative deviations in lattice parameters and bond lengths, typically within a few percent, are expected given the intrinsic differences between experimental and theoretical conditions. Experiments probe finite crystallites that may contain defects, residual functional groups, and environmental effects, whereas the present simulations describe ideal, defect-free crystals at zero temperature under periodic boundary conditions, using a nonlocal vdW functional. These comparisons support the view that the structural model adopted here captures the essential features of experimentally realized AA-stacked HsGDY while providing atomistic detail on bonding, stacking registry, and interlayer spacing that is difficult to extract directly from measurements.

\subsection{Energy Landscape}

The energetic stability of layered HsGDY was evaluated through cohesive and interlayer binding energies, which provide complementary measures of bulk thermodynamic robustness and stacking-driven stabilization. The cohesive energy per atom is defined as:

\begin{equation}
E_{\mathrm{coh}} = \frac{E_{\mathrm{bulk}} - \sum_{i} N_{i} E_{i}^{\mathrm{isol}}}{N_{\mathrm{tot}}},
\end{equation}

where $E_{\mathrm{bulk}}$ is the total energy of the crystalline phase, $E_{i}^{\mathrm{isol}}$ is the energy of an isolated atom of species $i$, $N_{i}$ is the number of atoms of type $i$, and $N_{\mathrm{tot}}$ is the total number of atoms. The interlayer binding energy per atom is obtained from:

\begin{equation}
E_{\mathrm{bind}} = \frac{E_{\mathrm{stack}} - N_{\mathrm{layer}} E_{\mathrm{mono}}}{N_{\mathrm{tot}}},
\end{equation}

where $E_{\mathrm{stack}}$ is the total energy of the stacked structure, $E_{\mathrm{mono}}$ is the energy of a single isolated layer constrained to the same in-plane geometry, and $N_{\mathrm{layer}}$ is the number of layers in the stacking sequence.

\begin{figure}[pos = h]
    \centering
    \includegraphics[width=\linewidth]{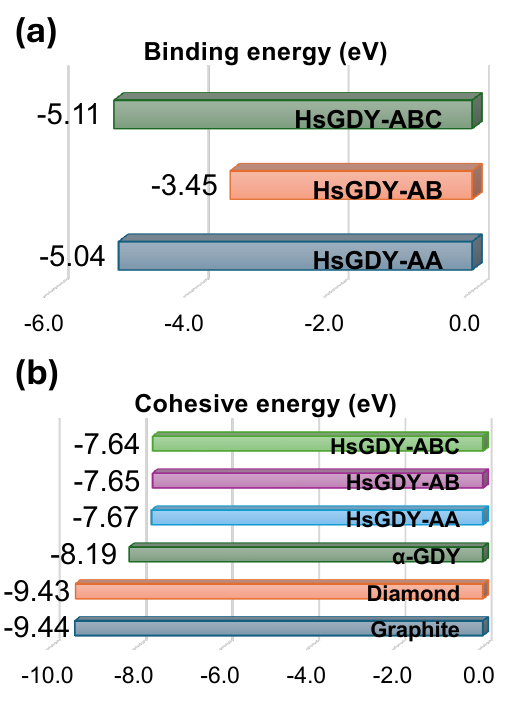}
    \caption{HsGDY Energies. (a) Binding energy per atom for AA, AB, and ABC stacking sequences, highlighting the near degeneracy between AA and ABC and the reduced stability of the AB registry. (b) Cohesive energy per atom for graphite, diamond, $\alpha$-GDY, and HsGDY, positioning the hydrogenated framework within the stability spectrum of carbon allotropes.}
    \label{fig:cohesive}
\end{figure}

To contextualize the thermodynamic stability of HsGDY, Figure~\ref{fig:cohesive} compares its cohesive energy with those of $\alpha$-GDY, graphite, and diamond, while also presenting the binding energies associated with distinct stacking registries.

The calculated cohesive energies follow the trend graphite $<$ diamond $<$ $\alpha$-GDY $<$ HsGDY, with values of $-9.44$, $-9.43$, $-8.19$, and $-7.67$ eV/atom, respectively. The smaller values observed for the graphdiyne-derived frameworks reflect their porous architectures and mixed sp-sp$^{2}$ bonding, which lead to lower atomic packing densities than the dense sp$^{2}$ and sp$^{3}$ networks of graphite and diamond. 

A direct comparison between $\alpha$-GDY and HsGDY reveals that hydrogen incorporation decreases the cohesive energy by approximately $0.5$ eV/atom. This decrease originates from the partial replacement of extended C-C $\pi$ conjugation by localized C-H $\sigma$-bonds, weakening the overall bonding network while preserving structural integrity. Such behavior aligns with previous studies on hydrogen-functionalized graphdiyne and related porous carbon frameworks.

The binding energy values indicate that AA and ABC configurations are nearly degenerate, with values of $-5.04$ and $-5.11$ eV/atom, whereas the AB stacking is markedly less stable at $-3.45$ eV/atom. The small energy difference between AA and ABC suggests that HsGDY accommodates multiple interlayer arrangements with comparable stabilization, pointing to a relatively shallow energy landscape. By contrast, the energetic cost associated with AB stacking highlights the strong dependence of interlayer cohesion on lateral registry.

This behavior emerges from the interplay between dispersive $\pi$-$\pi$ interactions and steric effects introduced by the out-of-plane C-H groups. In AA and ABC configurations, the relative alignment between pores and hydrogenated sites promotes a more favorable balance between attractive dispersion forces and short-range repulsion. The AB arrangement, however, yields suboptimal overlap patterns, thereby weakening interlayer coupling. These findings indicate that hydrogen functionalization not only modifies local bonding but also reshapes the interlayer energy landscape governing the 3D stackings.

The present results are consistent with the experimental realization of AA-stacked HsGDY reported by Liu \textit{et al.} \cite{liu2025aa}, whose combined experimental and computational study first established this configuration as the preferred stacking sequence. Our calculations further support this stacking as most stable, while extending the analysis to alternative registries within a fully periodic framework, thereby providing a broader atomistic perspective on the interlayer energy landscape. Minor quantitative variations in the relative stacking energies are expected, given the different theoretical approaches and structural models employed. Importantly, both studies conclude that AA stacking is the most favorable configuration, highlighting the stabilizing role of ordered pore channels, together with the interplay between dispersive $\pi$-$\pi$ interactions and C-H-mediated steric effects, in governing the assembly of the layered architecture.

\subsection{Dynamical and Thermal Stability}

\begin{figure}[pos = b!]
    \centering
    \includegraphics[width=\linewidth]{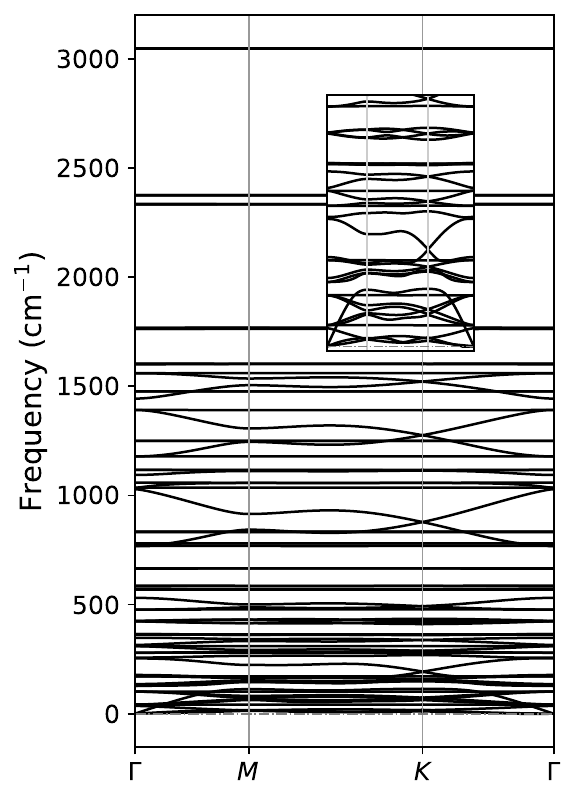}
    \caption{Phononic band structure of bulk HsGDY.}
    \label{fig:phonons}
\end{figure}

The structural robustness of layered HsGDY was further assessed through lattice dynamics and finite-temperature analyses. Figure~\ref{fig:phonons} presents the phonon dispersion of bulk HsGDY along the principal high-symmetry directions of the Brillouin zone.

The complete absence of imaginary frequencies demonstrates that HsGDY corresponds to a true minimum of the potential-energy surface and is therefore dynamically stable. The low-frequency region is dominated by acoustic and soft optical modes associated with interlayer shear and layer-breathing motions, indicating that the stacked architecture does not exhibit latent long-wavelength instabilities.

At higher frequencies, nearly dispersionless bands emerge as a characteristic feature of the vibrational spectrum. These flat modes originate predominantly from localized C-H vibrations together with stretching modes of the acetylenic C-C units, whose large force constants produce weakly dispersive phonons \cite{jenkins2024thd}. Such spectral separation provides clear vibrational fingerprints and reflects the coexistence of hydrogen functional groups with the rigid carbon backbone.

Thermal stability was further verified through AIMD simulations at 700 K. Throughout the simulation, the total energy fluctuated around approximately -126.819 eV/atom without a systematic drift, and no bond-breaking or bond-formation events were observed. The absence of structural degradation at this temperature indicates that the hydrogenated framework preserves its integrity under significant thermal excitation.

Taken together, the phonon spectrum and AIMD results consistently demonstrate that the HsGDY structures investigated here are both dynamically and thermally stable. These findings reinforce the energy landscape trends discussed previously and establish the layered hydrogen-substituted graphdiyne as a structurally robust 3D carbon framework suitable for applications requiring mechanical resilience and thermal reliability.

\subsection{Electronic Structure}

\begin{figure}[pos = h]
    \centering
    \includegraphics[width=\linewidth]{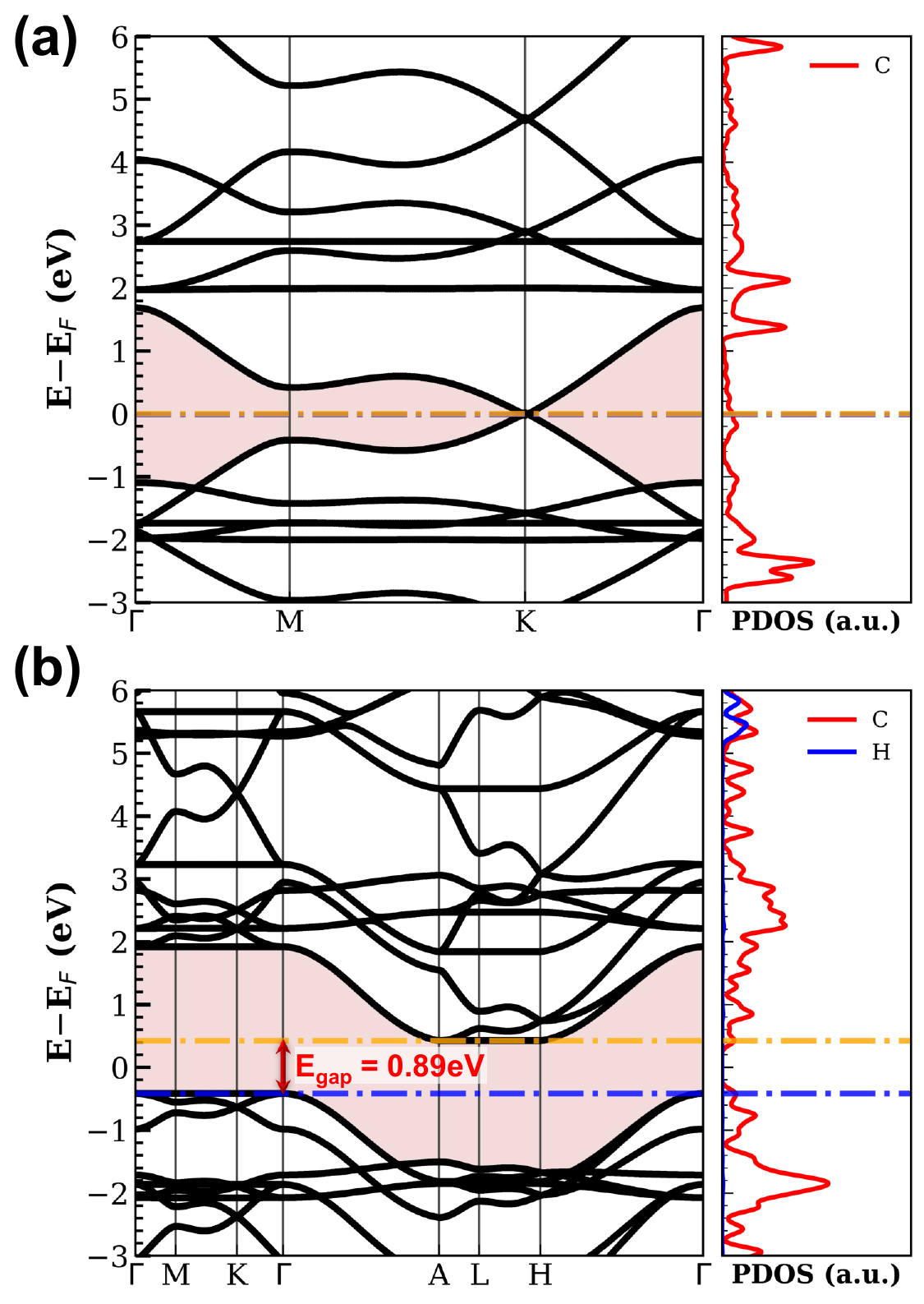}
    \caption{Electronic band structures and the corresponding projected density of states of: (a) monolayer $\alpha$-GDY and: (b) AA-stacked bulk HsGDY. The shaded region highlights the band gap, while the PDOS identifies the orbital contributions near the band edges.}
    \label{fig:bands-pdos}
\end{figure}

The electronic band structures and the corresponding projected density of states (PDOS) of $\alpha$-GDY and HsGDY are presented in Figure~\ref{fig:bands-pdos}, which directly shows how hydrogen substitution and interlayer stacking reshape the low-energy electronic landscape of the graphdiyne framework.

For $\alpha$-GDY, the bands intersect at the Fermi level with a Dirac-like dispersion previously reported in the literature \cite{Huang2018,Niu2013}. These linearly dispersive states originate predominantly from the C $2p_z$ orbitals belonging to the conjugated sp-sp$^2$ carbon network, confirming the semimetallic character associated with extended $\pi$ delocalization.

In contrast, AA-stacked HsGDY exhibits an indirect electronic band gap of approximately $0.89$ eV, evidencing the electronic reconfiguration induced by hydrogen functionalization. The formation of C-H bonds partially interrupts the extended $\pi$ system, lifting the Dirac degeneracy and stabilizing a semiconducting ground state. Notably, the magnitude of this gap remains smaller when compared with fully hydrogenated graphene-derived materials, which typically display gaps between $1$ and $4$ eV \cite{Haberer2010,Sofo2007,Son2016}, and with hydrogenated porous graphene networks that often reach values near $3.0$-$3.5$ eV \cite{Fabris2017,Fabris2018}. Functionalized graphdiyne monolayers may exhibit even larger gaps, approaching 5 eV, under extensive hydrogenation or halogenation \cite{Koo2014}. The comparatively smaller gap found in HsGDY therefore indicates that hydrogen incorporation creates spatially localized structural distortions and preserves a substantial fraction of the carbon $\pi$ backbone, yielding a moderate-gap semiconductor rather than a strongly localized sp$^3$-dominated framework.

The PDOS further clarifies the orbital origin of the frontier states. In both systems, the band edges are overwhelmingly dominated by C $2p_z$ contributions, demonstrating that the low-energy electronic response is governed by the carbon $\pi$ manifold. Hydrogen introduces mainly $\sigma$-type states and plays a secondary role near the Fermi level, with its spectral weight becoming more visible at deeper valence and higher conduction energies. This behavior indicates that hydrogen acts primarily as an electronic perturbation of the carbon network, modifying conjugation without creating an independent carrier channel.

Additional insights emerge when different stacking registries are compared. While AA-stacked HsGDY displays a gap of $0.89$ eV, the gap increases to $1.68$ eV and $1.89$ eV for the AB and ABC arrangements, respectively. This systematic gap widening reflects the sensitivity of the electronic structure to interlayer registry. Departures from the highly symmetric AA configuration reduce $\pi$-$\pi$ overlap between adjacent layers, weaken band dispersion near the Fermi level, and shift the valence-band maximum and conduction-band minimum further apart in energy. Such behavior demonstrates that stacking provides an additional degree of freedom for tuning the electronic response of HsGDY beyond chemical functionalization alone.

These results reveal a clear transition from the Dirac semimetallicity of $\alpha$-GDY to a stacking-dependent semiconducting regime in HsGDY. The coexistence of moderate band gaps with preserved $\pi$ connectivity suggests a favorable balance between carrier transport and electronic controllability, positioning this hydrogen-substituted framework as a promising platform for optoelectronic and energy-related applications.

\subsection{Optical properties}

\begin{figure}[pos = t!]
    \centering
    \includegraphics[width=\linewidth]{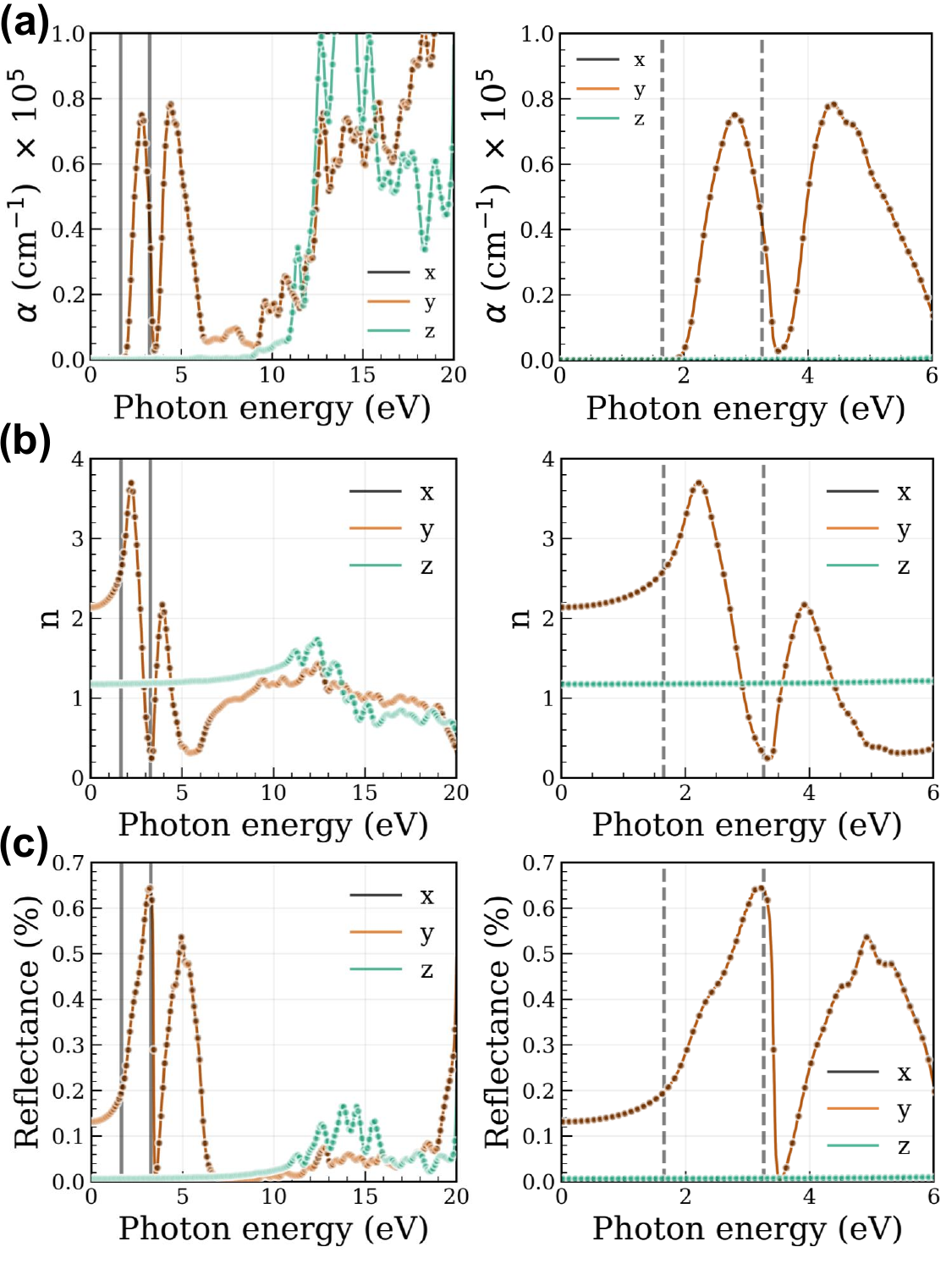}
    \caption{Optical absorption coefficient, refractive index, and reflectance of HsGDY for the 30-atom unit cell along the $x$, $y$, and $z$ directions. Dashed vertical lines indicate characteristic transition energies.}
    \label{fig:optical-unitcell}
\end{figure}

The anisotropic optical response of HsGDY was evaluated through the frequency-dependent absorption coefficient, refractive index, and reflectance along the crystallographic $x$, $y$, and $z$ directions. Figure~\ref{fig:optical-unitcell} presents the spectra obtained for the AA-HsGDY, adopted here as the representative description of the bulk optical behavior.

The absorption spectra reveal a pronounced optical anisotropy. The in-plane components ($x$ and $y$) exhibit intense features spanning the visible to near-UV regions, whereas the out-of-plane response remains significantly weaker throughout the investigated photon-energy range. This behavior directly reflects the layered architecture and the predominance of in-plane $\pi$ conjugation, which enhances dipole-allowed transitions for light polarized parallel to the carbon backbone while suppressing vertical excitations.

The first strong absorption peaks emerge slightly above the fundamental electronic band gap, indicating that they originate primarily from direct interband transitions between frontier $\pi$ states. At higher photon energies, additional structures arise from transitions involving more dispersive $\pi$ and $\sigma$ bands, signaling the progressive participation of deeper valence states and higher conduction bands. This spectral evolution is consistent with the previously discussed moderate band gap and confirms that hydrogen substitution preserves an electronically active carbon network rather than producing a strongly localized sp$^{3}$ framework.

The refractive index values follow the same directional trend. Larger static values are obtained for in-plane polarization, while the $z$ component remains comparatively low, suggesting reduced optical confinement perpendicular to the layers. Reflectance remains moderate at low energies and increases near the main absorption peaks, indicating efficient light-matter coupling without the large metallic reflectivity typical of gapless systems. Collectively, these characteristics position HsGDY as a directional semiconductor with strong polarization selectivity.

To evaluate finite-size effects, the optical response was also computed for a 240-atom supercell that preserves the AA-stacked topology while extending the real-space periodicity of the porous framework. The spectral profiles remain essentially unchanged, with only minor energy shifts and slight intensity redistributions. No additional low-energy transitions appear, demonstrating that the primitive cell already captures the relevant local electronic environment and that artificial confinement effects are negligible at the scales considered.

These results establish that the optical behavior of bulk HsGDY is primarily governed by its anisotropic $\pi$ network and stacking-preserved electronic structure. The coexistence of a moderate electronic band gap, strong in-plane absorption, and limited out-of-plane response suggests that hydrogen-substituted graphdiyne may serve as a promising platform for polarization-sensitive photodetectors, anisotropic optoelectronic components, and layered photonic architectures.

\section{Conclusion and perspectives}

This work presents a comprehensive first-principles investigation of 3D-HsGDY, establishing clear relationships between stacking configuration, structural stability, and emergent physicochemical properties. Structural analyses demonstrate that hydrogenation preserves the hexagonal $sp$–$sp^{2}$ carbon framework while expanding the in-plane lattice and introducing localized out-of-plane distortions at hydrogenated sites. The resulting layered crystals exhibit well-defined interlayer separations on the order of $3.3$–$3.4$ $\r{A}$, and the energetic landscape reveals that AA and ABC stackings are nearly degenerate and energetically favored over the AB arrangement, indicating that HsGDY can accommodate distinct interlayer registries without compromising structural integrity.

Dynamical stability is confirmed by the complete absence of imaginary phonon modes across the Brillouin zone, while high-frequency flat bands are associated with C–H vibrations and stretching modes of the acetylenic C–C units. Thermal robustness is further supported by AIMD simulations at 700 K, which maintain an average energy throughout the trajectory and show no observable bond breaking or formation. Together, these results demonstrate that the investigated structures are both dynamically and thermally stable.

From an electronic perspective, hydrogenation drives a transition from the Dirac-like semimetallic behavior of $\alpha$-GDY to a moderate-gap semiconductor. AA-stacked HsGDY exhibits an indirect band gap of $0.89$ eV dominated by C $2p_{z}$ states, whereas alternative stackings further widen the gap, highlighting interlayer registry as an additional degree of freedom for band-gap tuning. The optical response is strongly anisotropic, with intense in-plane absorption extending from the visible to the near-ultraviolet region and a markedly weaker out-of-plane component, reflecting the layered topology and the persistence of in-plane $\pi$ conjugation.

The present results provide atomistic support to recent experimental advances on AA-stacked HsGDY and extend their interpretation by clarifying the energetic competition between stacking sequences and its consequences for electronic and optical behavior. By demonstrating the coexistence of structural robustness, stacking-tunable electronic properties, and pronounced optical anisotropy, this study positions HsGDY as a promising porous carbon framework for next-generation optoelectronic and energy-related technologies.

\section*{Acknowledgements}
Guilherme S. L. Fabris acknowledges the São Paulo Research Foundation (FAPESP) fellowship (process number $\#$2024/03413-9). Raphael B. de Oliveira thanks the National Council for Scientific and Technological Development (CNPq) (process numbers 151043/2024-8 and 200257/2025-0). Bruno Ipaves thanks CNPq process number \#194155/2025-0 and FAPESP process numbers \#2024/11016-0. Marcelo L. Pereira Junior acknowledges financial support from FAPDF (grant 00193-00001807/2023-16), CNPq (grants 444921/2024-9 and 308222/2025-3), and CAPES (grant 88887.005164/2024-00).
Douglas S. Galvão acknowledges the Center for Computing in Engineering and Sciences at Unicamp for financial support through the FAPESP/CEPID Grant (process number $\#$2013/08293-7). We thank the Coaraci Supercomputer Center for computer time (process number $\#$2019/17874-0). Douglas S. Galvao also acknowledges support from INEO/CNPq and FAPESP grant 2025/27044-5.

\printcredits
\bibliographystyle{unsrt}
\bibliography{references}

\end{document}